\documentclass[reprint,aps,prb,superscriptaddress,nobibnotes]{revtex4-1}%
\usepackage{amsmath}
\usepackage{amssymb}
\usepackage{natbib}
\usepackage{graphicx}
\usepackage{xcolor}
\usepackage{amsfonts}%
\providecommand{\U}[1]{\protect\rule{.1in}{.1in}}
\setcitestyle{numbers,square}
\begin{document}

\title{Magnon Trap by Chiral Spin Pumping}
\author{Tao Yu}\thanks{These authors contributed equally to this work.} 
\affiliation{Max Planck Institute for the Structure and Dynamics of Matter, Luruper Chaussee 149, 22761
	Hamburg, Germany}
\author{Hanchen Wang}
\thanks{These authors contributed equally to this work.} 
\affiliation{Fert Beijing Institute, School of Microelectronics, Beijing Advanced Innovation Center for Big Data and Brain Computing, Beihang University, Beijing 100191, China}
\author{Michael A.~Sentef}
\affiliation{Max Planck Institute for the Structure and Dynamics of Matter, Luruper Chaussee 149, 22761
	Hamburg, Germany}
\author{Haiming Yu} 
\affiliation{Fert Beijing Institute, School of Microelectronics, Beijing Advanced Innovation Center for Big Data and Brain Computing, Beihang University, Beijing 100191, China}
\author{Gerrit E. W. Bauer}
\affiliation{WPI-AIMR \&  Institute for Materials Research  \& CSRN, Tohoku
	University, Sendai 980-8577, Japan}
\date{\today}

\begin{abstract}
Chiral spin pumping is the generation of a unidirectional spin current in half
of ferromagnetic films or conductors by dynamic dipolar stray fields from
close-by nanomagnets. We formulate a general theory of long-range chiral
interactions between magnets mediated by unidirectional traveling waves, e.g.,
spin waves in a magnetic film or microwaves in a waveguide. The traveling
waves emitted by an excited magnet can be perfectly trapped by a second,
initially passive, magnet by a dynamical interference effect. When both
magnets are excited by a uniform microwave, the chiral interaction between
them creates a large imbalance in their magnon numbers.

\end{abstract}
\maketitle

\section{Introduction}

Unidirectional propagation of quasiparticles is a fundamental phenomenon with
practical interest for information processing in logic devices
\cite{nano_optics,near_field,Petersen,poineering_1,poineering_2,Springer_book}%
. Magnons, the elementary excitations of the magnetic order, carry an
intrinsic angular momentum that can be utilized to transport information
\cite{magnonics1,magnonics2,magnonics3,magnonics4}. Dynamic dipolar stray
fields emitted by ferromagnetic nanostructures can generate a unidirectional
magnon current in a ferromagnetic film or conductors in its proximity by
``chiral spin pumping"
\cite{chiral_simulation,Yu1,Yu2,chiral_pumping,electron}. Magnons can
propagate over centimeters \cite{centimeter} in magnetic insulators such as
yttrium iron garnet (YIG) without Joule heating. In contrast to electrons that
are easily controlled and confined by electric gates, the electric control of
magnons on a small length scale is difficult. Magnons are trapped by
inhomogeneous magnetic fields in, e.g., a spin-polarized atomic hydrogen gas
\cite{cold_atom1} and superfluid $^{3}$He-B \cite{cold_atom2,cold_atom3}.
Existing magnon transistors \cite{transistor,magnon_valve} do not fully trap
magnons in the film because inefficient gating.

\begin{figure}[th]
\begin{center}
{\includegraphics[width=8.8cm]{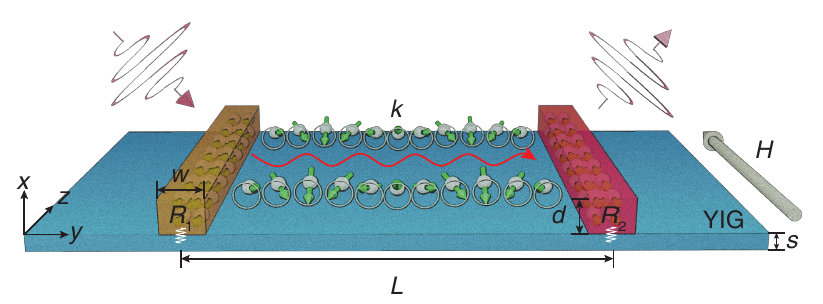}}
\end{center}
\caption{Two magnetic nanowires on top of a thin YIG film with in-plane
magnetization along the magnetic field $H\hat{\mathbf{z}}$. A local stripline
antenna (not shown) is used to excite and detect the magnetization dynamics in
the nanowires. The red arrow indicates the unidirectional magnon current in
the film. The geometric parameters used in the text are indicated.}%
\label{configuration}%
\end{figure}

In this work, we theoretically demonstrate trapping of waves on short length
scales by the unique functionalities of chiral pumping
\cite{chiral_pumping,waveguide,Petersen,poineering_1}. We first focus on a
device consisting of two magnetic transducers in the form of nanowires on top
of a high-quality thin film of a magnetic insulator such as YIG
(Fig.~\ref{configuration}). Exciting one of the nanomagnets by external
microwaves launches spin waves in the magnetic film that propagate in one
direction only \cite{chiral_simulation,Yu1,Yu2,chiral_pumping}. These spin
waves then interact with the second nanowire that does not see the microwaves
directly and excite its magnetization, which then in turn emits spin waves as
well. The relative phase shift of the magnetizations in the two wires is
$\pi+\phi_{k}$, where $\phi_{k}$ is the transmission phase of the spin waves
in the film. The phase shift $\pi$ is caused by twice the dissipative phase
shift at the resonance of two identical nanowires \cite{phase_shift}. When the spin waves from
both sources interfere destructively outside the two wires, the nanowires form
a magnonic cavity that confines the traveling spin waves irrespective of the
geometric phase $\phi_{k}$ caused by their distance $L$. The spin waves
thereby cannot escape the passive wire, they are trapped. Since spin waves are
not reflected back and forth to form standing waves, this mechanism is robust
with respect to disorder and implies nearly perfect spin and energy transfer
between the wires. The entrapment of traveling waves by the dynamic $\pi$
phase shift may occur in other chirally coupled systems as well: Two magnets
located on a special line of a waveguide at which the momentum and rotation
direction are locked \cite{waveguide} can trap the photons in the same manner
\cite{Canming_trap}.

This paper is organized as follows. We first introduce the general trapping
mechanism in Sec.~\ref{general_mechanism} and then discuss two typical cases
including magnetic wires on top of a magnetic film in Sec.~\ref{case_1} and
magnetic spheres in a microwave waveguide in Sec.~\ref{case_2}. An imbalanced
pumping between two magnets by chiral interaction is addressed in
Sec.~\ref{imbalanced_pumping}. We conclude with Sec.~\ref{discussion}.

\section{General trapping mechanism}

\label{general_mechanism} Generally, we consider a chirally coupled system in
one dimension, in which the magnon mode in the $l$-th magnet $\hat{\beta}_{l}$
of frequency $\omega_{\mathrm{K},l}$ at position $R_{l}\hat{\mathbf{y}}$
interacts with the traveling modes $\hat{\alpha}_{k}$ of frequency $\omega
_{k}$, e.g., in Fig.~\ref{configuration} the magnon modes are the Kittel modes
\cite{Kittel} in the nanowires and the traveling modes are spin waves in the
magnetic film \cite{chiral_simulation,Yu1,Yu2,chiral_pumping}. Another typical
example we shall highlight is two magnetic spheres in a microwave waveguide in
which the magnon modes are the Kittel modes in the spheres and the traveling
waves are the microwaves
\cite{waveguide,Canming_trap,chiral_waveguide1,chiral_waveguide2,chiral_waveguide3}%
. The model is extendable to chiral nanooptics \cite{nano_optics} and
plasmonics \cite{Petersen} in which the magnetic dipoles by magnon modes may
be replaced by electrical dipoles \cite{electric_dipole1,electric_dipole2}.
The chiral coupling, with mechanism addressed later, implies that the magnon
modes prefer to interact with the traveling waves propagating in one
direction. We use a general quantum description between harmonic oscillators
that is allowed to generally describe the chirally coupled systems. We denote
the coupling constant between the magnon mode and traveling waves as $g_{k,l}$
and the full chiral coupling indicates that one of $g_{|k|}$ and $g_{-|k|}$
vanishes. The general Hamiltonian then reads \cite{chiral_pumping}
\begin{align}
\hat{H}/\hbar &  =\sum_{l}\omega_{\mathrm{K},l}\hat{\beta}_{l}^{\dagger}%
\hat{\beta}_{l}+\sum_{k}\omega_{k}\hat{\alpha}_{k}^{\dagger}\hat{\alpha}%
_{k}\nonumber\\
&  +\sum_{l}\sum_{k}\left(  g_{k,l}e^{-ikR_{l}}\hat{\beta}_{l}\hat{\alpha}%
_{k}^{\dagger}+g_{k,l}e^{ikR_{l}}\hat{\beta}_{l}^{\dagger}\hat{\alpha}%
_{k}\right)  .
\end{align}

We now consider two identical magnets located at $\mathbf{r}_{1}=R_{1}%
\hat{\mathbf{y}}$ and $\mathbf{r}_{2}=R_{2}\hat{\mathbf{y}}$, which act as
transducers for microwaves that are emitted or detected by local microwave
antennas. They communicate by exciting and absorbing the traveling waves.
Hereafter, $g_{k,1}=g_{k,2}=g_{k}$ and $\omega_{\mathrm{K},1}=\omega
_{\mathrm{K},2}=\omega_{\mathrm{K}}$. Expressing the local magnon operators at
$R_{1}$ and $R_{2}$ by $\hat{\beta}_{1}$ and $\hat{\beta}_{2}$, we obtain the
long-range chiral interaction between the two magnets mediated by the
traveling waves through eliminating the dynamics of traveling modes in the
equations of motion of the system \cite{input_output1,input_output2} (see
Appendix~\ref{App_A} for derivation):
\[
\frac{d}{dt}\left(
\begin{matrix}
\hat{\beta}_{1}\\
\hat{\beta}_{2}%
\end{matrix}
\right)  +i\left(
\begin{matrix}
\tilde{\omega}_{\mathrm{K}}-i\Gamma(\omega) & -i\Gamma_{12}(\omega)\\
-i\Gamma_{21}(\omega) & \tilde{\omega}_{\mathrm{K}}-i\Gamma(\omega)
\end{matrix}
\right)  \left(
\begin{matrix}
\hat{\beta}_{1}\\
\hat{\beta}_{2}%
\end{matrix}
\right)  =\left(
\begin{matrix}
\hat{P}_{1}\\
\hat{P}_{2}%
\end{matrix}
\right)  .
\]
Here, $\tilde{\omega_{\mathrm{K}}}=\omega_{\mathrm{K}}-i\kappa/2$, in which
$\kappa=2\alpha_{\mathrm{G}}\omega_{\mathrm{K}}$ is the intrinsic damping of
the Kittel modes in the magnets parameterized by the Gilbert coefficient
$\alpha_{\mathrm{G}}$. $\hat{P}_{l}\equiv-\sqrt{\kappa_{p}^{(l)}}\hat
{p}_{\mathrm{in}}^{(l)}$ represent the input terms from the local antennas
$\hat{p}_{\mathrm{in}}^{(l)}$ to the Kittel magnons, where $\kappa_{p}^{(l)}$
is the additional radiative damping induced by the microwave photons that is
usually much smaller than $\kappa$. With $R_{2}>R_{1}$ in mind, the couplings
between magnets read
\begin{align}
\Gamma_{12}(\omega)  &  =\frac{1}{v(k_{\omega})}|g_{-k_{\omega}}%
|^{2}e^{ik_{\omega}(R_{2}-R_{1})},\nonumber\label{dissipative_1}\\
\Gamma_{21}(\omega)  &  =\frac{1}{v(k_{\omega})}|g_{k_{\omega}}|^{2}%
e^{ik_{\omega}(R_{2}-R_{1})},
\end{align}
and the self-interaction
\begin{equation}
\Gamma(\omega)=\frac{1}{2v(k_{\omega})}\left(  |g_{k_{\omega}}|^{2}%
+|g_{-k_{\omega}}|^{2}\right)  \label{addition_damping1}%
\end{equation}
is the pumping-induced damping for a single nanowire \cite{chiral_pumping}.
Here, $v(k)$ is the group velocity of the traveling waves and $k_{\omega}$ is
the positive root of $\omega_{k}=\omega$, and we have used the on-shell
approximation $\omega=\omega_{\mathrm{K}}$ at the ferromagnetic resonance
(FMR). $|\Gamma_{12}(\omega)|\neq|\Gamma_{21}(\omega)|$ since $|g_{k}%
|\neq|g_{-k}|$, implying the (partially) chiral dissipative coupling
\cite{dissipative1,dissipative2,dissipative3,Canming}. In the fully chiral
limit with, e.g., $g_{-k}=0$, $|\Gamma_{21}(\omega)|=2\Gamma(\omega)$, i.e.,
twice the magnon broadening by chiral pumping Eq.~(\ref{addition_damping1}).
When one of the couplings is exactly zero, one magnet can influence the other
magnet but without back action. This breaks the reciprocity of the interaction
and promises new functionalities as addressed below.

We now only turn on $\hat{p}_{\mathrm{in}}^{(1)}$ and calculate the excited
traveling waves. In frequency space and the chiral limit we have (see
Appendix~\ref{App_A})
\begin{align}
\hat{\alpha}_{k}(\omega)  &  =G_{k}\left(  \omega\right)  g_{k} \left(
e^{-ikR_{1}}\hat{\beta}_{1}(\omega)+e^{-ikR_{2}}\hat{\beta}_{2}(\omega
)\right)  ,\nonumber\\
\hat{\beta}_{2}(\omega)  &  =\frac{-i\sum_{k}g^{2}_{k}G_{k}\left(
\omega\right)  e^{ik(R_{2}-R_{1})}}{-i(\omega-\omega_{\mathrm{K}}
)+\kappa/2+i\sum_{k}g_{k}^{2}G_{k}\left(  \omega\right)  }\hat{\beta}%
_{1}(\omega), \label{phase_relation}%
\end{align}
where $G_{k}\left(  \omega\right)  =1/\left[  (\omega-\omega_{k})+i\kappa
_{k}/2\right]  $ is the Green's function of the traveling modes, where
$\kappa_{k}$ denotes the intrinsic damping of traveling modes.
Equation~(\ref{phase_relation}) gives the phase relation between $\hat{\beta
}_{2}$ and $\hat{\beta}_{1}$ when the left magnet is excited. At the FMR,
\begin{equation}
\hat{\beta}_{2}(\omega_{\mathrm{K}})=\eta(\omega_{\mathrm{K}})e^{i\pi
+ik_{r}(R_{2}-R_{1})}\hat{\beta}_{1}(\omega_{\mathrm{K}}),
\label{phase_relation2}%
\end{equation}
where $k_{r}$ is the positive root of $\omega_{k_{r}}=\omega_{\mathrm{K}}$,
and
\begin{equation}
\eta(\omega_{\mathrm{K}})=\frac{2\Gamma(\omega_{\mathrm{K}}) }{\kappa
/2+\Gamma(\omega_{\mathrm{K}}) } \label{coefficient}%
\end{equation}
modulates the magnitude of the excited magnon amplitude. This corresponds to a
phase shift
\begin{align}
\Delta\phi=\pi+k_{r}(R_{2}-R_{1})
\end{align}
between the two magnets. $k_{r}(R_{2}-R_{1})$ is the phase delay by the
traveling wave transmission between the two magnets. The phase shift of $\pi$
reflects the doubled dissipative phase shifts $\pi/2$ between magnons in the
magnets and traveling waves that is the key for the magnon trap addressed
below. We have recently reported observation of this phase shift with two
magnetic nanowires on top of magnetic film by microwave spectroscopy
\cite{Haiming}. Remarkably, when $\kappa/2\ll\Gamma$, $\eta\rightarrow2$,
implying that the energy accumulates in the passive magnet, apparently
amplifying the signal by a factor of 2.

The phase relation Eq.~(\ref{phase_relation2}) implies the trapping of magnons at
the FMR when $\eta(\omega_{\mathrm{K}})\rightarrow1$, i.e., when the
pumping-induced damping and the intrinsic damping are comparable. At the FMR,
the excited traveling-wave amplitude with momentum $k_{r}$ reads
\begin{equation}
\left\langle \hat{\alpha}_{k_{r}}(\omega_{\mathrm{K}})\right\rangle =G_{k_{r}%
}\left(  \omega_{\mathrm{K}}\right)  g_{k_{r}}e^{-ik_{r}R_{1}}\left\langle
\hat{\beta}_{1}(\omega_{\mathrm{K}})\right\rangle \left(  1-\eta
(\omega_{\mathrm{K}})\right)  ,\label{keys}%
\end{equation}
which indicates suppression of the right-propagating waves on the right side
of the magnets when $\eta(\omega_{\mathrm{K}})\rightarrow1$. Meanwhile, the
left-propagating waves are not excited due to the nature of chiral coupling.
Therefore, the excited traveling waves are confined between magnets and the
magnons are trapped in the right magnet (the spatial amplitude is calculated
below). By tuning $\eta$ one can modulate the transport of traveling waves as well.

The above mechanism is universal in chirally coupled harmonic oscillators. We
address two typical examples in optomagnonics below including coupled magnetic
wires and film \cite{Yu1,Yu2,chiral_pumping,Haiming} and coupled magnetic
spheres and microwave waveguide
\cite{waveguide,chiral_waveguide1,chiral_waveguide2,chiral_waveguide3}.

\section{Magnetic nanowire and film}

\label{case_1} We consider the effectively one-dimensional model in
Fig.~\ref{configuration} with two sufficiently long magnetic nanowires
(thickness $d$ and width $w$) on top of a thin YIG film of thickness $s$. The
latter is of the order of tens of nanometers, such that the excited
magnetization is distributed uniformly across the film without chirality
itself \cite{Yu1,chiral_pumping}. The distance between the nanowires is $L\gg
w$. Magnons in the nanowires are excited and detected by local metal stripline
antennas on top of the nanowires \cite{Haiming}. The interlayer exchange
interaction between the wire and film has been found to be smaller than the
dipolar one in the antiparallel configuration \cite{Yu1,Yu2}, and can be
further suppressed by a spacer without affecting the longer-range dipolar coupling.

We focus on the linear regime at temperatures far below the critical one. To
leading order, the magnetization operators in the magnetic wires and film may
be expanded by \cite{Kittel,HP}
\begin{align}
&  \hat{M}_{\alpha}(\mathbf{r})=-\sqrt{2M_{s}\gamma\hbar}\sum_{k}\left(
m_{\alpha}^{(k)}(x)e^{iky}\hat{\alpha}_{k}+\mathrm{H.c.}\right)  ,\nonumber\\
&  \hat{\tilde{M}}_{\alpha,l}(\mathbf{r})=-\sqrt{2\tilde{M}_{s,l}\gamma\hbar
}\left(  \tilde{m}_{l,\alpha}^{\mathrm{K}}(\mathbf{r})\hat{\beta}%
_{l}+\mathrm{H.c.}\right)  , \label{expansion}%
\end{align}
where $M_{s}$ and $\tilde{M}_{s,l}$ are the saturated magnetizations of film
and nanowire, $-|\gamma|$ is the electron gyromagnetic ratio, $m_{\alpha
}^{(k)}(x)$ and $\tilde{m}_{l,\alpha}^{\mathrm{K}}(\mathbf{r})$ represent,
respectively, the amplitudes of the spin waves in the film and Kittel modes in
the wires, and $k$ denotes $k_{y}$. The magnetization $\mathbf{M}$ in the film
couples to the dipolar field emitted by the magnetization $\tilde{\mathbf{M}%
}_{l}$ of the wire via the Zeeman interaction \cite{Landau,chiral_pumping}.
With the mode expansion, the coupling constants (refer to
Appendix~\ref{App_B1})
\begin{equation}
g_{k,l}=-F_{l}(k)\left(  m_{x}^{(k)\ast},m_{y}^{(k)\ast}\right)  \left(
\begin{array}
[c]{cc}%
|k| & ik\\
ik & -|k|
\end{array}
\right)  \left(
\begin{array}
[c]{c}%
\tilde{m}_{l,x}^{\mathrm{K}}\\
\tilde{m}_{l,y}^{\mathrm{K}}%
\end{array}
\right)
\end{equation}
are real and the form factor
\begin{equation}
F_{l}(k)=\frac{2\mu_{0}\gamma}{k^{3}}\sqrt{\frac{M_{s}\tilde{M}_{s,l}%
}{\mathcal{L}}}(1-e^{-|k|d})(1-e^{-|k|s})\sin\left(  \frac{kw}{2}\right)  .
\end{equation}
$\mathcal{L}$ is the (sufficiently large) length of the magnetic nanowire. By
tuning the magnetic field to change the resonant momentum $k$ of the spin
waves to the Kittel mode, the factor $\sin(kw/2)$ allows for tuning of the
dipolar coupling strength. Chiral coupling is reflected by $g_{-|k|}=0$ for
the circularly polarized spin waves with $m_{y}^{(k)}=im_{x}^{(k)}$
($\mathbf{M}\parallel\hat{\mathbf{z}}$) \cite{chiral_pumping}. In the
Appendix~\ref{App_C}, the full numerical simulation confirms the chiral spin
pumping and validates the single-mode approximation.

We now calculate the excited magnetization in real space. By Eq.~(\ref{keys}),
at the FMR $\omega\rightarrow\omega_{\mathrm{K}}$ the film magnetization in
Eq.~(\ref{expansion}) is the real part of
\begin{align}
&  \hat{M}_{\alpha}(x,y)=-2\sqrt{2M_{s}\gamma\hbar}\hat{\beta}_{1}%
(\omega_{\mathrm{K}})\sum_{k}m_{\alpha}^{(k)}(x)G_{k}(\omega_{\mathrm{K}%
})g_{k}\nonumber\\
&  \times\left(  e^{-ik(R_{1}-y)}-\eta(\omega_{\mathrm{K}})e^{ik_{r}%
(R_{2}-R_{1})}e^{-ik(R_{2}-y)}\right)  ,
\end{align}
in which the $k$-integral can be carried out by closing the contour in the
complex plane, with singularities in the denominator of $G_{k}$ being $k_{\pm
}^{\ast}=\pm(k_{r}+i\epsilon)$, where $\epsilon$ is the inverse of the
propagation decay length. When $y<R_{1}<R_{2}$, the integral
path is chosen in the lower half plane that selects the singularity $k_{-}%
^{\ast}$, leading to
\begin{align}
\hat{M}^{L}_{\alpha}(x)  &  =\frac{2i}{v_{k_{r}}}\sqrt{2M_{s}\gamma\hbar}%
\hat{\beta}_{1}(\omega_{\mathrm{K}})m_{\alpha}^{(k_{r})}(x)g_{-k_{r}%
}\nonumber\\
&  \times\left(  e^{ik_{r}(R_{1}-y)}-\eta(\omega_{\mathrm{K}})e^{ik_{r}%
	(2R_{2}-R_{1}-y)}\right)  ,
\end{align}
which vanishes when the chiral coupling $g_{-k}=0$. 
 When
$y>R_{2}>R_{1}$, the integral path is chosen in the upper half plane that
selects the singularity $k_{+}^{\ast}$, and we obtain
\begin{align}
\hat{M}_{\alpha}^{R}(x)  &  =\frac{2i}{v_{k_{r}}}\sqrt{2M_{s}\gamma\hbar}%
\hat{\beta}_{1}(\omega_{\mathrm{K}})m_{\alpha}^{(k_{r})}(x)g_{k_{r}%
}\nonumber\\
&  \times e^{-ik_{+}^{\ast}(R_{1}-y)}\left(  1-\eta(\omega_{\mathrm{K}%
})\right)  ,
\end{align}
which vanishes when $\eta(\omega_{\mathrm{K}})\rightarrow1$. When
$R_{1}<y<R_{2}$,
\begin{equation}
\hat{M}_{\alpha}^{M}(x)=\frac{2i}{v_{k_{r}}}\sqrt{2M_{s}\gamma\hbar}\hat
{\beta}_{1}(\omega_{\mathrm{K}})m_{\alpha}^{(k_{r})}(x)g_{k_{r}}%
e^{-ik_{+}^{\ast}(R_{1}-y)}%
\end{equation}
is a right-propagating wave in the chiral limit. We note that the decay of the
excited magnetization is governed by the ubiquitous Gilbert damping by the
complex $k_{+}^{\ast}$. Without going into the details of the non-chiral
system, we envision that the vanishing magnetization on the right side of the
passive nanowire is also established even without chiral coupling. But the
left-moving traveling waves are excited and the chiral spin pumping can even
emerge in the non-chiral system with active and passive excitations by the
dynamic interference effect. We note that the trapped magnetization is not a
standing wave as there are no back and forth reflections. This helps to focus
the magnetization to a small region of micrometers and efficiently transport
the spin information directly from one wire to the other.

\begin{figure}[th]
\begin{center}
{\includegraphics[width=7.5cm]{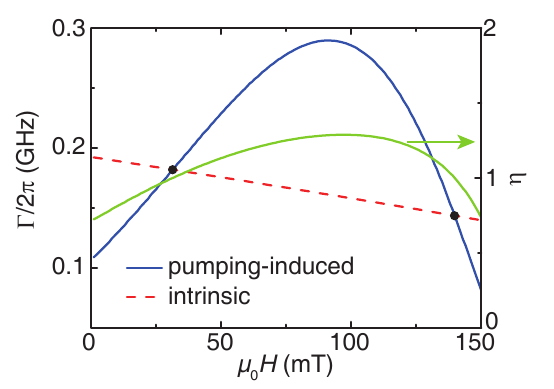}}
\end{center}
\caption{Pumping-induced $\left(  \Gamma\right)  $ and intrinsic $\left(
\kappa/2\right)  $ broadenings of a Kittel magnon as function of applied
magnetic field. At the crossings (blue dots) the magnons emitted from the left
nanowire are perfectly trapped. The material parameters are given in the text.
}%
\label{damping}%
\end{figure}

We illustrate the concept by calculating the pumping-induced damping and
magnon trapping under chiral pumping of spin waves for Co nanowires of
thickness 30~nm and width 100~nm on top of a YIG film with $s=20$~nm. We use
the magnetizations $\mu_{0}M_{s}=0.177$~T for YIG and $\mu_{0}\tilde{M}%
_{s}=1.62$~T for Co \cite{Yu2}. The intrinsic Gilbert damping coefficient of
Co wire is taken to be $\alpha_{\mathrm{G}}=0.01$
\cite{Yu2,Co_damping,Haiming}. Figure~\ref{damping} is the plot of the
magnetic-field dependence of $\eta$, the pumping-induced broadening $\Gamma$
and intrinsic one $\kappa/2=\alpha_{\mathrm{G}}\omega_{\mathrm{K}}$ of the
wire Kittel mode, which can be measured in terms of the broadening of the wire
FMR. For particular magnetic fields $\mu_{0}H\approx31.8$ and 139.7~mT, the
pumping-induced damping equals the intrinsic one, at which the trapping
becomes perfect. 

\begin{figure}[th]
\begin{center}
{\includegraphics[width=8.8cm]{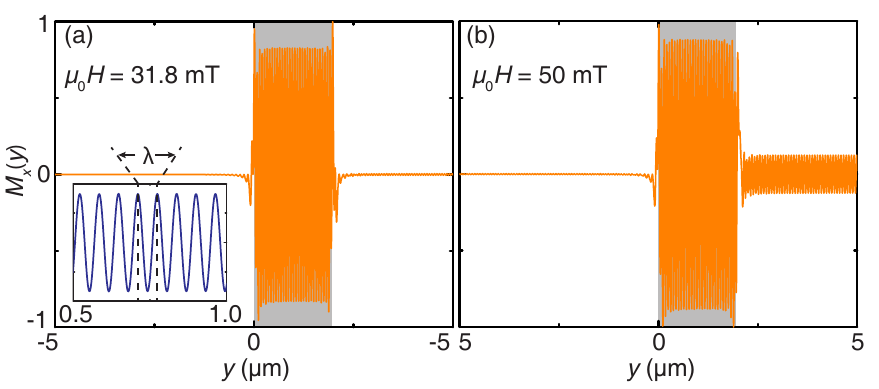}}
\end{center}
\caption{Snapshots of the calculated magnetization $\mathbf{M}_{x}$ in real
space for magnetic fields $\mu_{0}H\approx31.8$ [(a)] and 50~mT [(b)]. The
material parameters are given in the text. }%
\label{mx_spatial}%
\end{figure}

In Figs.~\ref{mx_spatial}(a) and (b) we plot a snapshot of $\mathbf{M}_{x}$ in
real space for magnetic fields $\mu_{0}H\approx31.8$ and 50~mT. We choose Co
wires centered at $R_{1}=0$ and $R_{2}=2~\mathrm{\mu}$m. At the critical field
31.8~mT, the excited magnetization is very well confined between the two wires
[(a)], while magnetization is allowed to leak into the right half-space
otherwise [(b)]. This device therefore functions as a magnon
valve/switch/transistor that can be opened and closed by weak magnetic fields
with characteristics far superior to previous realizations that operate by
very different principles \cite{transistor,magnon_valve}. 

\section{Magnetic spheres and microwave waveguide}

\label{case_2} Another example in optomagnonics is the chirally coupled
magnetic spheres and microwaves in a waveguide along the $\hat{\mathbf{y}}$
direction with a rectangular cross section of dimensions $a>b$ ($a\parallel
\hat{\mathbf{z}}$, $b\parallel\hat{\mathbf{x}}$)
\cite{waveguide,chiral_waveguide1,chiral_waveguide2,chiral_waveguide3}.
Considering the lowest TE$_{10}$ mode with magnetic-field component
$\mathcal{H}_{k,x}=0$, the magnetic field is polarization-momentum locked at
the special positions termed \textquotedblleft chiral line" with the magnetic
field $\mathcal{H}_{k,z}=i\mathrm{sgn}(k)\mathcal{H}_{k,y}$ (details are shown
in Appendix~\ref{App_B2}). Two identical sub-mm magnetic spheres with
saturation magnetization $M_{s}$ and volume $V_{s}$ are saturated in the
$\hat{\mathbf{x}}$-direction and put on the chiral lines ${R}_{l=1,2}%
\hat{\mathbf{y}}$. The diameter of the spheres is much smaller than the
wavelength of the microwaves such that they can be treated as point particles
\cite{Weichao_simulation}. The magnetization and waveguide magnetic field are
quantized by (see Appendix~\ref{App_B2})
\begin{align}
\hat{\tilde{M}}_{\alpha,l}(\mathbf{r})  &  =-\sqrt{2M_{s}\gamma\hbar}\left(
\tilde{m}_{l,\alpha}^{\mathrm{K}}(\mathbf{r})\hat{\beta}_{l}+\mathrm{H.c.}%
\right)  ,\nonumber\\
H_{\beta}(\mathbf{r})  &  =\sum_{k}\left(  \mathcal{H}_{\beta,k}%
(x,z)e^{iky}\hat{\alpha}_{k}+\mathrm{H.c.}\right)  ,
\end{align}
and the Zeeman coupling leads to the coupling constant
\begin{equation}
g_{k,l}=-\mu_{0}\sqrt{\frac{\gamma M_{s}V_{s}}{2\hbar}}e^{ikR_{l}}\left(
\mathcal{H}_{k,y}\left(  x,z\right)  -i\mathcal{H}_{k,z}\left(  x,z\right)
\right)  .
\end{equation}
At the chiral line, the magnet only couples with the microwave propagating
along one direction. The formalism is exactly the same as the one for magnetic
wires and film. Again, with condition $\eta=1$ satisfied, excitation of the
left active magnet by a local antenna can confine the microwaves between two
magnets and the photon is trapped at the passive magnet: the microwaves at the
right of the passive magnet
\begin{equation}
\hat{H}_{\alpha}=-\frac{2i}{v_{k_{r}}}\hat{\beta}_{1}(\omega_{\mathrm{K}%
})\mathcal{H}_{\alpha}^{(k_{r})}g_{k_{r}}e^{-ik_{+}^{\ast}(R_{1}-y)}\left(
1-\eta(\omega_{\mathrm{K}})\right)
\end{equation}
vanish and the microwave is unidirectionally invisible \cite{Canming_trap}.
This proves that the dynamic interference effect is an universal mechanism for trapping.

\section{Imbalanced pumping}

\label{imbalanced_pumping} We turn to the situation in which the whole sample
is illuminated by a global microwave field with finite $\hat{p}_{\mathrm{in}%
}^{(1)}$ and $\hat{p}_{\mathrm{in}}^{(2)}$. With the same coherent driving, we
may expect that chirality causes different magnon populations in the two
wires. We set up the master equation of the density operator $\hat{\rho}$ to
calculate the dynamics driven by the microwaves, $\hat{P}_{l}(t)=\hat{P}%
_{l}(0)(e^{-i\omega_{d}t}+e^{i\omega_{d}t})$ with frequency $\omega_{d}$. In
the chiral limit $\Gamma_{12}=0$ and at the FMR, the master equation in the
rotating frame becomes \cite{master_equation} (refer to Appendix~\ref{App_A}
for construction)
\begin{align}
\partial_{t}\hat{\rho}  &  =i\left[  \hat{\rho},\sum_{l=1,2}\Omega_{l}%
(\hat{\beta}_{l}+\hat{\beta}_{l}^{\dagger})\right]  +\left[  \hat{\rho}%
,\frac{\Gamma_{21}}{2}\hat{\beta}_{2}^{\dagger}\hat{\beta}_{1}-\frac
{\Gamma_{21}^{\ast}}{2}\hat{\beta}_{1}^{\dagger}\hat{\beta}_{2}\right]
\nonumber\\
&  +\sum_{l}\Gamma_{l}\hat{\mathcal{L}}_{ll}\hat{\rho}+\frac{\Gamma_{21}%
^{\ast}}{2}\hat{\mathcal{L}}_{12}\hat{\rho}+\frac{\Gamma_{21}}{2}%
\hat{\mathcal{L}}_{21}\hat{\rho},
\end{align}
where $\Omega_{l}=i\langle\hat{P}_{l}(0)\rangle$ represents the drive
amplitude and $\mathcal{L}_{ij}\hat{\rho}=2\hat{\beta}_{j}\hat{\rho}\hat
{\beta}_{i}^{\dagger}-\hat{\beta}_{i}^{\dagger}\hat{\beta}_{j}\hat{\rho}%
-\hat{\rho}\hat{\beta}_{i}^{\dagger}\hat{\beta}_{j}$ is the Lindblad
superoperator that accounts for the relaxation. Denoting the average of an
operator as $\langle\hat{Q}(t)\rangle=\langle\hat{Q}\rho(t)\rangle$, the
driven magnon amplitudes in the steady state become
\begin{align}
&  \langle\hat{\beta}_{1}\rangle={\langle\hat{P}_{1}\rangle}/{\tilde{\Gamma}%
},\nonumber\\
&  \langle\hat{\beta}_{2}\rangle={\langle\hat{P}_{2}\rangle}/{\tilde{\Gamma}%
}-{\Gamma_{21}\langle\hat{P}_{1}\rangle}/{\tilde{\Gamma}^{2}},
\label{amplitude}%
\end{align}
where $\tilde{\Gamma}=\kappa/2+\Gamma$. When the excitation microwave is
uniform with the same amplitude $\langle\hat{P}_{1}\rangle=\langle\hat{P}%
_{2}\rangle=i\Omega$ and identical magnetic nanowires with small intrinsic
damping $|\tilde{\Gamma}|\rightarrow|\Gamma_{21}|/2$. The ratio of the
steady-state magnon populations in the two magnetic nanowires
\begin{equation}
{\langle\hat{\beta}_{2}^{\dagger}\hat{\beta}_{2}\rangle}/{\langle\hat{\beta
}_{1}^{\dagger}\hat{\beta}_{1}\rangle}=5-4\cos\left(  k_{r}(R_{2}%
-R_{1})\right)  ,
\end{equation}
and hence can be on the order of ten and is tunable over a wide range by
changing their separation or $k_{r}$ by the Kittel frequency. This
amplification is caused by the chiral dissipative coupling between magnets,
through which one magnet can input energy to another without back action. This
effect can be enhanced by adding more magnets \cite{waveguide}.

\section{Discussions}

\label{discussion} In conclusion, we propose a method to control spin wave
transport by weak magnetic fields based on the theory of chiral pumping of
spin waves. By exploiting two nanowires that communicate by unidirectional
spin waves, we achieve new functionalities such as magnon trapping,
amplification and a valve/transistor effect. The spin pumping by active and
passive magnets is different from conventional situation as it gives quite
different behavior of pumped current. The spatial distribution of magnons can
be detected inductively via microwave emission of a third magnetic wire
(supposing weak disturbance on the magnonic cavity) \cite{Yu2}, NV center
magnetometry \cite{Toeno}, Brillouin light scattering \cite{BLS}, and
electrically by the inverse spin Hall effect with a normal metallic wire such
as Pt \cite{van_Wees}. Replacing the nanowires by other objects such as
magnetic spheres or qubits, and the unidirectional spin waves by other
propagating quasiparticles such as waveguide photons, surface plasmons,
electrons or phonons, we envision our mechanism to be extended to other fields
including optomagnonics, nanooptics \cite{nano_optics}, quantum optics,
plasmonics \cite{near_field,Petersen}, spintronics, and spin mechanics.

\vskip0.25cm \begin{acknowledgments}
T.~Y.~and M.~A.~S.~acknowledge funding through the DFG Emmy Noether program (SE 2558/2-1). H.~W.~and H.~Y.~are supported by NSF China under Grants No. 11674020 and No. U1801661. G.B. is supported by JSPS KAKENHI Grant No. 19H006450.
\end{acknowledgments}

\begin{appendix}

\section{Long-range chiral interactions}

\label{App_A} In this section, we derive the long-range chiral interaction
between remote magnets from the equations of motion, based on which we
construct an effective non-Hermitian Hamiltonian for setting up the master
equations used in the main text. From the total Hamiltonian [Eq.~(1) in the
main text], the equations of motion of the magnons $\hat{\beta}_{l}$ and
traveling modes $\hat{\alpha}_{k}$ read \cite{input_output1,input_output2}
\begin{align}
i\frac{d\hat{\beta}_{l}(t)}{dt} &  =\omega_{\mathrm{K}}\hat{\beta}_{l}%
(t)+\sum_{k}g_{k}e^{ikR_{l}}\hat{\alpha}_{k}(t)-i\frac{\kappa+\kappa_{p}%
^{(l)}}{2}\hat{\beta}_{l}(t)\nonumber\\
&  -i\sqrt{\kappa_{p}^{(l)}}\hat{p}_{\mathrm{in}}^{(l)}(t),\nonumber\\
i\frac{d\hat{\alpha}_{k}(t)}{dt} &  =\omega_{k}\hat{\alpha}_{k}(t)+g_{k}%
\sum_{l}e^{-ikR_{l}}\hat{\beta}_{l}(t)-i\frac{{\kappa_{k}}}{2}\hat{\alpha}%
_{k}(t).\label{EOMs}%
\end{align}
Here, $\kappa=2\alpha_{\mathrm{G}}\omega_{\mathrm{K}}$ is the intrinsic
damping of the Kittel modes in the magnets (e.g., the magnetic nanowire or the
magnetic spheres) parameterized by the Gilbert coefficient $\alpha
_{\mathrm{G}}$, $\kappa_{p}^{(l)}$ is the additional radiative damping induced
by the microwave photons $\hat{p}_{\mathrm{in}}^{(l)}$, i.e., the coupling of
the magnet with the microwave antennas, and $\kappa_{k}$ denotes the intrinsic
damping of the traveling waves with momentum $k$. Integrating the second
equation in Eq.~(\ref{EOMs}) yields
\begin{align}
\hat{\alpha}_{k}(t) &  =\hat{\alpha}_{k,\mathrm{in}}e^{-i(\omega_{k}%
-i\kappa_{k}/2)t}-\sum_{l}ig_{k}e^{-ikR_{l}}\nonumber\\
&  \times\int_{-\infty}^{t}d\tau\hat{\beta}_{l}(\tau)e^{-i(\omega_{k}%
-i\kappa_{k}/2)(t-\tau)},
\end{align}
where $\hat{\alpha}_{k,\mathrm{in}}$ is the input of the traveling waves that
can be set to be zero without direct excitation. This leads to
\begin{align}
\frac{d\hat{\beta}_{l}(t)}{dt} &  =-i\omega_{\mathrm{K}}\hat{\beta}%
_{l}(t)-\frac{\kappa}{2}\hat{\beta}_{l}(t)-\sum_{l^{\prime}}\sum_{k}%
|g_{k}|^{2}e^{ik(R_{l}-R_{l^{\prime}})}\nonumber\\
&  \times\int_{-\infty}^{t}\hat{\beta}_{l^{\prime}}(\tau)e^{-i(\omega
_{k}-i\kappa_{k}/2)(t-\tau)}+\hat{P}_{l}(t),\label{EOM}%
\end{align}
where $\hat{P}_{l}(t)\equiv-\sqrt{\kappa_{p}^{(l)}}\hat{p}_{\mathrm{in}}%
^{(l)}(t)$ and we have disregarded the microwave-induced dissipative damping
as usually $\kappa_{p}^{(l)}\ll\kappa$. The third term on the r.h.s of
Eq.~(\ref{EOM}) gives the effective interaction of magnons mediated by the
traveling waves. The magnons are assumed to move coherently within Markov
approximation, i.e., $\hat{\beta}_{l}(\tau)=\hat{\beta}_{l}(t)e^{i\omega
(t-\tau)}$. Thus,
\begin{equation}
\int_{-\infty}^{t}\hat{\beta}_{l^{\prime}}(\tau)e^{-i(\omega_{k}-i\kappa
_{k}/2)(t-\tau)}\approx\hat{\beta}_{l^{\prime}}(t)\frac{i}{\omega-\omega
_{k}+i\kappa_{k}/2},
\end{equation}
and Eq.~(\ref{EOM}) becomes
\begin{align}
\frac{d\hat{\beta}_{l}(t)}{dt} &  =-i\omega_{\mathrm{K}}\hat{\beta}%
_{l}(t)-\frac{\kappa}{2}\hat{\beta}_{l}(t)-i\sum_{l^{\prime}}\sum
_{k}e^{ik(R_{l}-R_{l^{\prime}})}\nonumber\\
&  \times\frac{|g_{k}|^{2}}{\omega-\omega_{k}+i\kappa_{k}/2}\hat{\beta
}_{l^{\prime}}+\hat{P}_{l}(t)\nonumber\\
&  =-i\omega_{\mathrm{K}}\hat{\beta}_{l}(t)-\frac{\kappa}{2}\hat{\beta}%
_{l}(t)-\Gamma_{l}(\omega)\hat{\beta}_{l}(t)\nonumber\\
&  -\sum_{ll^{\prime}}\Gamma_{ll^{\prime}}(\omega)\hat{\beta}_{l^{\prime}%
}(t)+\hat{P}_{l}(t).\label{EOM_all}%
\end{align}
Here, we have defined the damping by pumping the traveling-wave ($l^{\prime
}=l$)
\begin{equation}
\Gamma_{l}(\omega)=i\sum_{k}\frac{|g_{k}|^{2}}{\omega-\omega_{k}+i\kappa
_{k}/2}=\frac{1}{2v(k_{\omega})}\left(  |g_{k_{\omega}}|^{2}+|g_{-k_{\omega}%
}|^{2}\right)  \label{addition_damping}%
\end{equation}
and the traveling-wave mediated effective interaction ($l\neq l^{\prime}$)
\begin{align}
\Gamma_{ll^{\prime}}(\omega) &  =i\sum_{k}e^{ik(R_{l}-R_{l^{\prime}})}%
\frac{|g_{k}|^{2}}{\omega-\omega_{k}+i\kappa_{k}/2}\nonumber\\
&  =\left\{
\begin{matrix}
\frac{1}{v(k_{\omega})}|g_{k_{\omega}}|^{2}e^{ik_{\omega}|R_{l}-R_{l^{\prime}%
}|},\quad~\mathrm{when}~~R_{l}>R_{l^{\prime}}\\
\frac{1}{v(k_{\omega})}|g_{-k_{\omega}}|^{2}e^{ik_{\omega}|R_{l}-R_{l^{\prime
}}|},\quad\mathrm{when}~~R_{l}<R_{l^{\prime}}%
\end{matrix}
\right.  ,\label{coupling}%
\end{align}
where $v(k)$ is the group velocity of the traveling waves and $k_{\omega}$ is
the positive root of $\omega_{k}=\omega$. In Eqs.~(\ref{addition_damping}) and
(\ref{coupling}), we have assumed $\kappa_{k}\rightarrow0_{+}$ by assuming a
high quality of, e.g., magnetic film or microwave waveguide. With two
identical magnets $R_{2}>R_{1}$, the Heisenberg equation of motion
Eq.~(\ref{EOM_all}) is recovered when we define the effective non-Hermitian
Hamiltonian
\begin{equation}
\hat{H}_{\mathrm{eff}}=\sum_{l=1}^{2}(\omega_{\mathrm{K}}-i\kappa
/2-i\Gamma_{l})\hat{\beta}_{l}^{\dagger}\hat{\beta}_{l}-i\Gamma_{12}\hat
{\beta}_{1}^{\dagger}\hat{\beta}_{2}-i\Gamma_{21}\hat{\beta}_{2}^{\dagger}%
\hat{\beta}_{1}.
\end{equation}
The chiral dynamics is then governed by a non-Hermitian Hamiltonian, which may
be separated into Hermitian $\hat{H}_{h}$ and anti-Hermitian $\hat{H}_{nh}$
parts as
\begin{align}
\hat{H}_{h} &  =\frac{\hat{H}_{\mathrm{eff}}+\hat{H}_{\mathrm{eff}}^{\dagger}%
}{2}\nonumber\\
&  =\sum_{l=1}^{2}\omega_{\mathrm{K}}\hat{\beta}_{l}^{\dagger}\hat{\beta}%
_{l}+i\frac{\Gamma_{21}^{\ast}-\Gamma_{12}}{2}\hat{\beta}_{1}^{\dagger}%
\hat{\beta}_{2}+i\frac{\Gamma_{12}^{\ast}-\Gamma_{21}}{2}\hat{\beta}_{1}%
\hat{\beta}_{2}^{\dagger},\nonumber\\
\hat{H}_{nh} &  =\frac{\hat{H}_{\mathrm{eff}}-\hat{H}_{\mathrm{eff}}^{\dagger
}}{2}\nonumber\\
&  =-i\sum_{l=2}^{2}\tilde{\Gamma}\hat{\beta}_{l}^{\dagger}\hat{\beta}%
_{l}-i\frac{\Gamma_{12}+\Gamma_{21}^{\ast}}{2}\hat{\beta}_{1}^{\dagger}%
\hat{\beta}_{2}-i\frac{\Gamma_{12}^{\ast}+\Gamma_{21}}{2}\hat{\beta}%
_{1}^{\dagger}\hat{\beta}_{2},
\end{align}
where $\tilde{\Gamma}=\kappa/2+\Gamma$. The coupling between magnons has both
coherent and dissipative components. The dissipative coupling between magnons
is responsible for the collective damping. The master equation used in the
main text is constructed based on $\hat{H}_{h}$ and $\hat{H}_{nh}$.

\section{Chirally coupled system}

\subsection{Coupled magnetic wire and film}

\label{App_B1} The first chirally coupled system we shall address is the
dipolarly coupled magnetic nanowire and film \cite{Yu1,Yu2,chiral_pumping}. We
assume the nanowire magnetization is along the wire $-\hat{\mathbf{z}}%
$-direction, in antiparallel to the film one. The Fourier components of the
dipolar field generated by a circularly polarized Kittel mode in the wire are
chiral. Considering a nanowire of thickness $d$ and width $w$, the magnetic
fluctuations are the real part of
\begin{align}
\tilde{M}_{x,y}(\mathbf{r},t)  &  =\tilde{m}_{x,y}\Theta(x)\Theta
(-x+d)\Theta(y+w/2)\Theta(-y+w/2)\nonumber\\
&  \times e^{-i\omega t},
\end{align}
where $\Theta(x)$ is the Heaviside step function. The magnon amplitudes read
\cite{chiral_pumping}
\begin{equation}
\tilde{m}_{x}=\sqrt{\frac{1}{4\mathcal{D}wd}}, ~~~~\tilde{m}_{y}=-i\sqrt
{\frac{\mathcal{D}}{4wd}}, \label{nanowire_waves}%
\end{equation}
where, by the applied magnetic field $H_{\mathrm{app}}$ and demagnetization
factors $N_{xx}\simeq w/(d+w)$ and $N_{yy}=d/(d+w)$ \cite{Yu1},
\begin{equation}
\mathcal{D}=\sqrt{\frac{H_{\mathrm{app}}+N_{xx}\tilde{M} _{s}}{H_{\mathrm{app}%
}+N_{yy}\tilde{M}_{s}}}.
\end{equation}
The corresponding dipolar magnetic field
\begin{equation}
\tilde{h}_{\beta}(\mathbf{r},t)=\frac{1}{4\pi}\partial_{\beta}\partial
_{\alpha}\int\frac{\tilde{M}_{\alpha}(\mathbf{r}^{\prime},t)}{|\mathbf{r}
-\mathbf{r}^{\prime}|}d\mathbf{r}^{\prime},
\end{equation}
and below the nanowire ($x<0$) the Fourier components
\begin{align}
\left(
\begin{array}
[c]{c}%
\tilde{h}_{x}(k,x,t)\\
\tilde{h}_{y}(k,x,t)
\end{array}
\right)   &  =-\frac{i}{4\pi}e^{\left\vert k\right\vert x}(1-e^{-\left\vert
k\right\vert d})\frac{2\sin(kw/2)}{k\left\vert k\right\vert }\nonumber\\
&  \times\left(
\begin{array}
[c]{cc}%
\left\vert k\right\vert  & ik\\
ik & -\left\vert k\right\vert
\end{array}
\right)  \left(
\begin{array}
[c]{c}%
\tilde{m}_{x}\\
\tilde{m}_{y}%
\end{array}
\right)  e^{-i\omega t}.
\end{align}
A perfectly left circularly polarized wire dynamics $\left(  \tilde{m}
_{y}=-i\tilde{m}_{x}\right)  $ implies that the Fourier components of
$\tilde{\mathbf{h}}$ with $k<0$ vanish. The Fourier component with $k>0$ is
perfectly right circularly polarized $\left(  \tilde{h}_{y}=i\tilde{h}%
_{x}\right)  $. However, a pure chiral coupling still arises even with
elliptically polarized Kittel mode as long as the spin waves in the film is
perfectly circularly polarized (see below). We assume the magnetic film of
thickness $s$ is sufficiently thin (tens of nanometer) such that the
dipolar-exchange spin waves are circularly polarized. With the film magnon
operator $\hat{\alpha}_{k}$ and amplitude ${m}^{(k)}_{x,y}$, the film
magnetization feel the dipolar field from the $l$-th nanowire centered at
$R_{l}$ via the Zeeman coupling
\begin{equation}
\hat{H}_{\mathrm{int}}=-\mu_{0}\int_{0}^{s}dxd\pmb{\bf \rho}M _{\beta
}(x,\pmb{\rho})\tilde{h}_{l,\beta}(x,\pmb{\rho}), \label{interaction}%
\end{equation}
leading to the coupling Hamiltonian between the magnons $\hat{\alpha}_{k}$ and
$\hat{\beta}_{l}$
\begin{equation}
\hat{H}_{c}=\hbar\sum_{l}\sum_{k}g_{l}(k)\hat{\beta}_{l}^{\dagger}\hat{\alpha
}_{k}+\mathrm{H.c.},
\end{equation}
with the coupling constant \cite{chiral_pumping}
\begin{align}
g_{l}(k)  &  =-2\mu_{0}\gamma\sqrt{\tilde{M}_{s}M_{s}}\frac{1}{k^{3}}%
\sin\left(  \frac{kw}{2}\right)  e^{ikR_{l}} \left(  1-e^{-|k|d}\right)
\nonumber\\
&  \times\left(  1-e^{-|k|s}\right)  \left(  m_{x}^{{(k)}*},m_{y}^{{(k)}%
*}\right)  \left(
\begin{array}
[c]{cc}%
|{k}| & ik\\
ik & -|k|
\end{array}
\right)  \left(
\begin{array}
[c]{c}%
\tilde{m}_{x}\\
\tilde{m}_{y}%
\end{array}
\right)  .
\end{align}
The Kittel mode in the nanowire couples with the spin waves with right
circular polarization ($m^{(k)}_{y}=im^{(k)}_{x}$) propagating perpendicular
to the nanowire with perfect chirality.

\subsection{Coupled magnetic sphere and microwave waveguide}

\label{App_B2} Another chiral system is the magnets in a microwave waveguide
along the $\hat{\mathbf{y}} $-direction \cite{waveguide}. Focusing on the
lowest $\left(  \mathrm{TE}_{10}\right)  $ mode of a rectangular waveguide
with dimensions $a>b$, the magnetic fields read
\begin{align}
\mathcal{H}_{k,x}  &  =0,\nonumber\\
\mathcal{H}_{k,y}  &  =-i\sqrt{\frac{2}{ab}}\frac{\sqrt{A_{k}}}{\mu_{0}%
\omega_{k}}\frac{\pi}{a}\cos\left(  \frac{\pi z}{a}\right)  ,\nonumber\\
\mathcal{H}_{k,z}  &  =-\mathrm{sgn}(k)\sqrt{\frac{2}{ab}}\frac{\sqrt{A_{k}}%
}{\mu_{0}\omega_{k}}\sqrt{\left(  \frac{\omega_{k}}{c}\right)  ^{2}-\left(
\frac{\pi}{a}\right)  ^{2}}\sin\left(  \frac{\pi z}{a}\right)  , \label{rec}%
\end{align}
where $A_{k}=\hbar\omega_{k}/(2\epsilon_{0})$. The sign of the $z$-component
of the magnetic field depends on the propagation direction. Particularly, the
magnetic field becomes circularly polarized when $|\mathcal{H}_{k,y}%
|=|\mathcal{H}_{k,z}|$, leading to the chiral line with positions $z_{0}$
determined by
\begin{equation}
\sqrt{\Big(\frac{\omega_{k}}{c}\Big)^{2}-\Big(\frac{\pi}{a}\Big)^{2}}%
\sin\Big(\frac{\pi z_{0}}{a}\Big)=\pm\frac{\pi}{a}\cos\Big(\frac{\pi z_{0}}%
{a}\Big).
\end{equation}
At the chiral line the polarization of the microwaves is locked to the
momentum. The waveguide is loaded with $N$ identical YIG spheres with
gyromagnetic ratio $-\gamma$, saturation magnetization $M_{s}$, and volume
$V_{s}$ at $\mathbf{r}_{l}=\pmb{\rho}+R_{l}\hat{\mathbf{y}}$ with
$l\in\{1,2,\cdots,N\}$. The sub-mm spheres are much smaller than the photon
wavelength of $\mathcal{O}\left(  \mathrm{cm}\right)  $, so they can be
treated as point particles. The static magnetic field $\mathbf{H}%
_{\mathrm{app}}=\left(  H_{\mathrm{app}},0,0\right)  $ is sufficiently strong
to saturate the magnetization in the $\hat{\mathbf{x}}$-direction. The photons
and magnons are coupled by the Zeeman interaction
\begin{align}
\hat{H}_{c}  &  =-\mu_{0}\int\mathbf{\hat{H}}(\mathbf{r})\cdot\mathbf{\hat{M}%
}(\mathbf{r})d\mathbf{r,}\nonumber\\
&  =\hbar\sum_{l}\sum_{k} g_{l}(k)\hat{p}_{k}\hat{\beta}_{l}^{\dagger
}+\mathrm{H.c.}, \label{coupling_photon}%
\end{align}
with the coupling constant
\begin{equation}
g_{l}(k)=-\mu_{0}\sqrt{\frac{\gamma M_{s}V_{s}}{2\hbar}}e^{ikR_{l}} \left[
\mathcal{H}_{k,y}\left(  \boldsymbol{\rho}\right)  -i\mathcal{H}_{k,z}\left(
\boldsymbol{\rho}\right)  \right]  ,
\end{equation}
which depends on the position of the magnetic particles. It is chiral with one
of $g_{|k|}$ and $g_{-|k|}$ vanishes when the magnets are put on the chiral lines.

\section{Micromagnetic simulation}

\label{App_C} 
\begin{figure}[th]
\centering
\includegraphics[width=7cm]{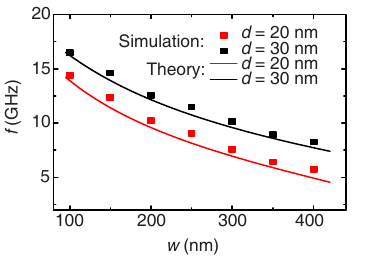} \caption{Simulated (red and black
dots) and analytical (red and black lines) results for the FMR frequencies of
a Co nanowire for thicknesses $d=20$ and 30~nm and widths $w$ from 100 to
400~nm.}%
\label{Comparison}%
\end{figure}

\begin{figure}[th]
\centering
\includegraphics[width=5.5cm]{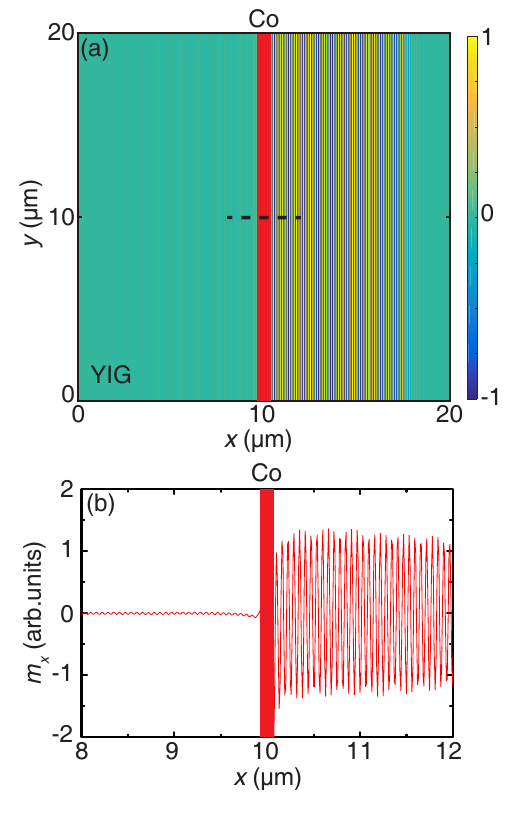} \caption{Simulation on the chiral
excitation by OOMMF (http://math.nist.gov/oommf). Parameters used for the
simulation are given in the text.}%
\label{map}%
\end{figure}

We carried out micromagnetic simulations by the public
object-oriented micromagnetic framework (OOMMF, http://math.nist.gov/oommf) in
order to justify the single-mode approximation in the magnetic nanowire and to
confirm that the excited spin waves flow into one direction in half of the
film. The dimensions of the YIG film is set to $20~\mathrm{\mu}$%
m$\times20~\mathrm{\mu}$m$\times20$ nm ($xyz$). A $100$~nm$\times
20~\mathrm{\mu}$m$\times20$~nm ($xyz$) Co nanowire is on the top of YIG at
$x=10~\mathrm{\mu}$m. The saturation magnetizations are 1200 kA/m and 140 kA/m
and the exchange constants are $13\times10^{-12}$ J/m and $3\times10^{-12}$
J/m for Co and YIG, respectively. The Gilbert damping of YIG is set to
$8\times10^{-5}$. We only include the interlayer dipolar interaction and set
the interlayer exchange interaction to zero. We created antiparallel
magnetization of Co and YIG by first applying a magnetic field of $-500$~mT to
saturate the magnetization of Co and YIG, followed by a $+50$~mT field that
switches only the YIG magnetization because the large shape anisotropy
stabilizes the Co magnetization. We excite the magnetic nanowire by a
monochromatic and uniform magnetic-field pulse that matches the mode frequency
of the magnetic nanowire that depends on the width $w$ and thickness $d$ via
the demagnetization factors $N_{yy}\simeq w/(d+w)$ and $N_{xx}\simeq d/(d+w)$
\cite{Yu1,Yu2,chiral_pumping},
\begin{equation}
\omega_{\mathrm{K}}=\mu_{0}\gamma\sqrt{(H_{\mathrm{app}}+N_{yy}\tilde{M}%
_{s})(H_{\mathrm{app}}+N_{xx}\tilde{M}_{s})}.
\end{equation}
The resonance frequencies of the Co nanowire are extracted by a fast Fourier
transform with 1000 time steps (10 ps) from the simulation. The simulated FMR
frequency in Fig.~\ref{Comparison} for isolated Co nanowires of thickness
$d=20$ and 30~nm and widths from $w=$100 to 400 nm agree with the above Kittel
formula. Next, we apply an oscillating field of $\mu_{0}H_{\mathrm{ext}%
}=0.2\times\sin(2\pi ft)$~mT with $f=15.1$~GHz to excite the FMR or a Co wire
on the top of YIG with $d=20$~nm and $w=100$~nm. The spatial map of the YIG
film magnetization $m_{x}$ is recorded after 1~ns as shown in Fig.~\ref{map},
where the red bar indicates the Co nanowire (top view). The lineplot for the
black dashed line ($x$ from 8 to 12~$\mathrm{\mu}$m) in Fig.~\ref{map}(a) is
shown in Fig.~\ref{map}(b). These results confirm the chiral excitation of
exchange spin waves by magnetodipolar coupling between the Co nanowire and YIG
film \cite{chiral_simulation}. At the FMR, the excited spin waves in figure have a single
wavelength that matches the frequency of the microwave pulse. This validates
the single-mode approximation used in our analytical treatment. An accurate
simulation of the magnon trapping by two wires cannot be done with the present
set-up. It requires a large device and a constant microwave drive and an
expensive effort that is beyond the scope of this work.
\end{appendix}

\end{document}